\documentstyle[11pt]{article}
\def\rmd{{\rm d}}

\newcommand{\qbarq}{\overline{q}q}
\newcommand{\uubar}{u \overline{u}}
\newcommand{\ssbar}{s \overline{s}}

\begin{document}

\begin{center}


\title
{The proton spin sum rule chiral bag prediction, an  update }

\vspace{8mm}

\author{
H.H\o gaasen \\
Fysisk Institutt, University of Oslo, \\
Blindern, N-0316 Oslo, Norway \\
and \\
F. Myhrer \\
Department of Physics and Astronomy,
 University of South Carolina, \\
Columbia, SC29208, USA }

\maketitle

\vspace{35mm}

{\bf Abstract.}

\end{center}

We reevaluate a quark model prediction using the new
QCD evolution function calculated to the 3 loop order and
conclude that this model compares favorably with  the new experimental results.

\newpage

{}From the semi-leptonic decays of baryons the matrix
element of the flavour changing parts of the axial currens are fairly well
known \cite{Gaillard,PDG}.
The first EMC measurements of the spin dependent proton structure
function $g_1^p(x,Q^2)$
when integrated, gave information of
the matrix element of the
flavour singlet part of the
axial current between proton states.
The first experimental results \cite{EMC1}
\begin{equation}
\Gamma_p(Q^2) = \int dx g_1^p(x,Q^2) =
0.126 \pm 0.012 (stat) \pm 0.026 (syst),
\label{oneold}
\end{equation}
\newline
implied that the matrix element of the  flavour singlet part of the
axial current between proton states was compatible with zero
within the error bars.
Despite the large experimental errors,
some models  of the baryon structure would give results
outside three standard deviations of the measured flavor singlet matrix
element value.
Since this was an unexpected result,
it generated a lot of discussion, see a short review in Ref. \cite{Theory},
and forced us to reconsider our present
model understanding of the baryonic structure.
However,
the Skyrmion model of Ref. \cite{EllisKa88} and the
 chiral bag model  \cite{HHFM1}
was immediately recognized to give
results that was within one standard deviation
of the measured integral over $g_1^p(x,Q^2)$.

\vspace{3mm}

In  the last two years there has been a lot of new important
experimental results coming from deep inelastic polarized
lepton scattering both on polarized proton, deuteron and helium targets
\cite{SMC,SMC93,E14293,Terrien94}.
It is of interest to see how
well these early model calculations compare
with the new experiments when we also take into
account recent progress in the calculation of QCD corrections.
The spin structure function $g_1^p(x,Q^2)$ has now been measured at quite
small values of x, and a smooth extrapolation to x=0
\footnote{See a recent discussion on this point by Close and Roberts
\cite{Close94}}
gives a new value for the integrated proton
spin structure function $\Gamma_p(Q^2)$ at
$\bar{Q}^2$ = 10 $GeV^2/c^2$ \cite{SMC}.
\footnote{
The SM Collaboration have
also evalated a "world average" \cite{SMC} at
$Q^2=10 GeV^2$ of
\begin{equation}
\Gamma_p(10 GeV^2) =0.142 \pm 0.008 \pm 0.011  )
\nonumber
\end{equation}
}
\begin{equation}
\Gamma_p(Q^2) =
0.136 \pm 0.011 \pm 0.011
\label{one}
\end{equation}
\noindent
This value is larger but compatible with the earlier result, eq.(\ref{oneold}).
To compare this new value with model prediction, requires a good
knowledge of QCD radiative corrections and
in particular the SU(3) flavor singlet
QCD evolution function, which, it was argued,
may strongly  affect  the quark model predictions \cite{Jaffe}.
Very recently this evolution function has been evaluated at the
3-loop level \cite{KvanN93},  and  confirmed by Larin
\cite{Larin}, who gives references to earlier works and who also
refers to the  order $\alpha^3$ QCD radiative
corrections calculations.

\vspace{3mm}

In the following we shall argue as if nature is SU(3) flavour symmetric,
not  because we believe that to be an extremely good approximation,
but because frequently this assumption is used in extracting
matrix elements of currents with a definite flavour from the experimental
semi-leptonic decay data \cite{Gaillard,PDG}.
Two main points are responsible for the present
model understanding of the baryon structure $\Gamma_p(Q^2)$.
The first is the role of the axial anomaly which we shall discuss
in the  chiral SU(3) limit.
The second point concerns the baryon wave function itself
and what we can learn from the other baryon spin observables.

\vspace{3mm}

To discuss the {\it first}  point we present our model picture of the
baryon structure.
We shall assume that
an effective  Lorentz-scalar force confines the quarks
to a finite region in space
(which we shall take to be a spherical cavity
in the rest system of the baryon) where they interact through the exchange
of gluons.
As the valence quarks are excluded from the QCD vacuum outside the cavity,
they will polarize the region just outside the cavity in $\qbarq$ pairs
in such a manner that
the axial current  carried by the valence quarks extends into
the QCD vacuum surrounding the cavity by  $\qbarq$ pairs
carrying the quantum numbers $J^P$ = $0^-$,
i.e., the pairs  are  the effective Goldstone bosons of the QCD vacuum.
For chiral symmetry to be a good symmetry, it is necessary
(but not sufficent) that the effective Goldstone fields in the QCD vacuum
couple to the axial currents of the quarks in the cavity such that
the axial current is continuous in all space.

\vspace{3mm}

In the chiral SU(3) limit the valence quarks (u,d,s) in the cavity
are massless.
The pseudoscalar Goldstone bosons surrounding the
quark confinement region, i.e.,
the pion, the  $\eta_8$ and the kaon, are also massless.
(This naturally imply that the $\qbarq$ content of the proton is mainly
a large distance phenomenon.)
On the other hand, the flavour singlet pseudoscalar, the $\eta^{\prime}$,
is massiv in the chiral limit due to the U(1) anomaly.
It is {\it this} consequence of the anomaly that we shall
explicitly use to illustrate how
a polarised  $\ssbar$ cloud could exist
around the  "bare" nucleon  (the nucleon's quark core),
even if  the "bare" nucleon contains no strange quarks.
To illustrate this we use the following simplistic picture:
Suppose for the sake of argument that we have a single confined u quark.
Since we are in the chiral limit  the axial current, which is
carried by the u-quark inside the confinement cavity,
has to be continued into the vacuum outside the cavity by massless
pseudoscalar mesons, which
would carry the flavour content of a $\uubar$ state.
This $\uubar$ state can be written in terms of states
that have well defined transformation properties
under SU(3) flavour at the
confinement "surface" where it changes into pseudoscalar mesons:

\begin{equation}
\uubar  = 1/\sqrt{2} \pi^0 + 1/\sqrt{3}  \eta^{\prime} +
1/\sqrt{6} \eta_8
\end{equation}

\noindent
If {\it all}  pseudoscalars were massless, the $\ssbar$ content of
$\eta_8$ and $\eta^{\prime}$ would cancel trivially,
as is implied by this equality.
However, the three pseudoscalar mesons have different masses, and
therefore their Yukawa ranges  beyond the cavity surface will be different.
For example, even in the chiral limit the
heavy $\eta^{\prime}$ (heavy due to the anomaly) and the
massless $\eta_8$ would generate a strangeness content in the proton
since the strength of the heavy $\eta^{\prime}$ field would
decay exponentially outside the confinement "surface".
Therefore, the  "bare" nucleon
containing only nonstrange quarks has a cloud of $\ssbar$ like states
even in the chiral SU(3) limit as a consequence of the U(1) anomaly.

\vspace{3mm}

The {\it second} $\,$ important point in these considerations is as follows:
If one wants to compute matrix
elements of the flavour singlet current, there is no reason to have confidence
in the results  if the model is not giving correct results for the matrix
elements of the flavour octet currents which are responsible for weak
semi-leptonic baryon decays.
{}From the experience with the MIT bag model \cite{MIT} we know that
relativistic quarks reduce the neutron axial charge
$g_A$ from 5/3 towards a more realistic value.
In addition, it is necessary to go beyond the
usual zero order SU(6)-like baryon wave functions
of an additive quark model.
An additive quark model gives
(in the chiral limit) for hyperon $\beta$- decays an
amplitude ratio, F/D = 2/3
like the non-relativistic quark model.
The fallacy of the additive quark models
becomes apparent for the baryon magnetic moments where
these quark models give the wrong inequality, i.e.,
for the measured magnetic moments
$|\mu(\Xi^-)| > |\mu(\Lambda)|$ \cite{PDG}.
These "spin" problems of $F$ and $D$ and of the baryon magnetic moments
are easily explained by
invoking the same  {\it effective} color magnetic spin-spin interaction that
give the "bare" nucleon - "bare" $\Delta$
mass splitting \cite{Ushio,Kobzarev,HHFM}.
This effective interaction introduces
dynamic spin-spin correlations in the baryon wave functions
that not only resolves the F/D problem, $F/D  \approx$  0.57, i.e.
$<$ 2/3, but also easily
explains why $| \mu(\Xi^-)|  >  | \mu(\Lambda)| $.
This spin-spin correlation contributes to the neutron's $g_A$ and
is important for the axial flavor-singlet charge \cite{HHFM1,MyTh}.

\vspace{3mm}

As should be evident from the above discussion,
to make a sound model prediction for $\Gamma_p(Q^2)$
and  $\Gamma_n(Q^2)$, it is
necessary that the quark model reproduces the value for
 $g_A (= 1.2573 \pm 0.0028)$
of the Bjorken sum rule as well as other baryon spin observables.
Then one can discuss the model prediction
for the flavor singlet part of the integrated proton spin structure
function and the so-called "spin content of the proton".
To be spesific we first define the relevant flavor SU(3)
amplitudes \cite{HHFM1,HHFM2}.
For $Q^2 \rightarrow \infty$
\begin{equation}
\Gamma_p(Q^2 \rightarrow \infty) =
< p\uparrow | \sum_i \frac{1}{2} \bar{\Psi_i} q_i^2
\gamma_5\gamma_3 \Psi_i | p \uparrow > \:
 = \frac{1}{6} [I_0 + I_8 + I_3 ]
\label{gamma1}
\end{equation}
where $q_i$ is the charge of the i'th quark.
The matrix element of the axial current operator
between proton states on the r.h.s. of eq.(\ref{gamma1})
is  evaluated at $Q^2$ = 0.
To make contact with the experimental result of eq.(\ref{one})
the axial current operator on the rigth-hand-side of eq.(\ref{gamma1})
should be renormalized at the point $Q^2$ = 10 $GeV^2/c^2$.
In eq.(\ref{gamma1}) the SU(3) flavor singlet amplitude is
\begin{equation}
I_0 = \sqrt{\frac{2}{3}}
< p \uparrow | \bar{\Psi} \gamma_5 \gamma_3 \lambda_0 \Psi | p \uparrow >
\label{I0}
\end{equation}
and the two flavor octet amplitudes are
\begin{equation}
I_8 = \frac{1}{2 \sqrt{3}}
< p\uparrow | \bar{\Psi} \gamma_5 \gamma_3 \lambda_8 \Psi | p \uparrow >
\label{I8}
\end{equation}
and
\begin{equation}
I_3 = \frac{1}{2}
< p\uparrow | \bar{\Psi} \gamma_5 \gamma_3 \lambda_3 \Psi | p\uparrow >
\label{I3}
\end{equation}
where $\lambda$ are the usual Gell-Mann matrices.
In terms of the standard SU(3) amplitudes $F$ and $D$ for the baryon
semi-leptonic decays, the two flavor octet amplitudes are

\vspace{3mm}

\hspace{3cm} $I_3 = \frac{1}{2} ( F + D )$ and $I_8 = \frac{1}{2} ( F - D/3 )$.

\vspace{3mm}

\noindent
The correspondence between the $I_i$'s and the commonly used
$\Delta q$'s are:
\vspace{3mm}

\hspace{3cm}  $\Delta u=\frac{1}{2} I_0 +I_8+I_3$,

\hspace{3cm}  $\Delta d=\frac{1}{2} I_0 +I_8-I_3$,

\hspace{3cm}  $\Delta s =\frac{1}{2} I_0 -2I_8$  ,

\vspace{2mm}

\noindent
and the "proton spin content" is:

\vspace{2mm}

\hspace{3cm}  $\Sigma=\Delta u+\Delta d+\Delta s =\frac{3}{2} I_0$

\vspace{3mm}

Since the two amplitudes $I_3$ and $I_8$ are matrix elements of conserved
currents in the chiral SU(3) limit, their values are
independent of the renormalization point.
Therefore, models
used to calculate $F$ and $D$ ($I_3$ and $I_8$)  should
have conserved axial octet currents
in the chiral limit in order to be reliably used at the scale of the SMC
experiment.
This is a key requirement in any model prediction of the
"proton spin content", a requirement which is  not satisfied by the
non-relativistic quark model, for example.
A concrete example of a model
satisfying the above  condition is the chiral bag model, an extention
of the MIT model.
Calculations within this model are quite straigtforward,
and give for many observables excellent insight  and
estimates of the relevant physics involved. This was (and is)
for us a valid reason for our use of this model.
One of the serious
drawback of this and similar models is the sharp confinement "wall"
(like a square well potential)
and the associated translational invariance problem.

\vspace{3mm}

Before presenting our quark model results
it is necessary to outline the radiative QCD corrections
to this sum rule and the possible renormalization of the flavor-singlet
axial current.
The proton sum rule including the QCD corrections at the scale
$Q^2$ have recently been calculated
at the 3-loop level of perturbative QCD (for references see \cite{Larin},
 and the later next order guestimates of Kataev \cite{Kataev})
and reads,
\begin{eqnarray}
\Gamma_p(Q^2) = \frac{1}{6} [ 1 - \frac{\alpha(Q^2)}{\pi}
 - 3.5833 \left( \frac{\alpha(Q^2)}{\pi} \right) ^2
 - 20.2153 \left( \frac{\alpha(Q^2)}{\pi} \right) ^3
- 130 \left( \frac{\alpha(Q^2)}{\pi} \right) ^4 ]\:
\nonumber    \\
(I_3 + I_8) +
\nonumber  \\
 + \frac{1}{6} [ 1 -  \frac{\alpha(Q^2)}{\pi}
 - 1.0959 \left( \frac{\alpha(Q^2)}{\pi} \right) ^2
- 3.7 \left( \frac{\alpha(Q^2)}{\pi} \right) ^3 ]\:
Y(\alpha(\mu^2),\alpha(Q^2))  I_0(\mu^2).
\label{QCD}
\end{eqnarray}
As discussed,
this expression assumes that the two flavour octet amplitudes, $I_3$ and
$I_8$,  are independent of the scale.
On the other hand the scale dependence of
the flavour singlet amplitude, $I_0(\mu^2)$, is made explicit in this
equation.
The effect of the anomalous dimensions of the flavor-singlet matrix
elements are taken care of by the
renormalization group (exponent) evolution function
$Y(\alpha(\mu^2),\alpha(Q^2))$, which
very recently has been evaluated at the 3-loop level
(in the lowest twist approximation),
e.g. \cite{Larin}, and  reads
\begin{equation}
ln Y(\alpha(\mu^2),\alpha(Q^2)) = \frac{2}{\pi}
\int_{\alpha(\mu^2)}^{\alpha(Q^2)}
d \alpha^{\prime} \: \frac{n_f - \gamma^{(1)} \frac{\alpha^{\prime}}{\pi}}
{\beta_0 + \frac{\beta_1}{4 \pi} \alpha^{\prime} + \frac{\beta_2}
{(4 \pi)^2} (\alpha^{\prime})^2}.
\label{Ymain}
\end{equation}
where $\gamma^{(1)}$ = (1 - $\frac{71}{12}$) $n_f$ + $\frac{1}{18}$
$n_f^2$ and $\beta_0$ = 11 - 2 $n_f$/3, $\beta_1$ = 102 - 38 $n_f$/3,
\newline
and
$\beta_2$ = $\frac{2857}{2}$ - $\frac{5033}{18} n_f$ +
$\frac{325}{54} n_f^2$.
The integrand is an increasing function of $\alpha '$ in the region
where the formula is valid.

\vspace{3mm}

To define the
renormalization group-invariant
(i.e., convention independent) nucleon matrix element of the singlet
axial current \cite{Larin,GluckReya}, we will write
the evolution function as a product
 $Y(\alpha(\mu^2),\alpha(Q^2))=Y(0,\alpha(Q^2)) \, Y(\alpha(\mu^2),0)$.
Then
\begin{equation}
 \Sigma_{inv} = exp \left( -  \frac{1}{4\pi}\int_{0}^{\alpha(\mu^2)}
\rmd  \alpha^{\prime}
\frac{\gamma(\alpha^{\prime})}{\beta(\alpha^{\prime})} \right)
\Sigma(\mu^2)=Y(\alpha(\mu^2),0) \Sigma(\mu^2),
\label{sigmainv}
\end{equation}

\noindent
and since $\alpha(Q^2)$ for $Q^2$ = 10 $GeV^2/c^2$ is small,
we can use eq.(\ref{Ymain}) and write the expansion for $n_f=3$
\begin{equation}
Y(0,\alpha(Q^2)) \approx  1 + 0.66667 \frac{\alpha(Q^2)}{\pi}
     + 1.213  \left( \frac{\alpha(Q^2)}{\pi} \right)^2 + \cdots
\label{Yapp1}
\end{equation}
\newpage
It then follows that the QCD corrections
for the integrated spin-structure function are contained in QCD
coefficients that are convention independent \cite{Larin,Kataev}:
\begin{eqnarray}
\Gamma_p(Q^2) = \frac{1}{6} [ 1 - \frac{\alpha(Q^2)}{\pi}
 - 3.5833 \left( \frac{\alpha(Q^2)}{\pi} \right) ^2
 - 20.2153 \left( \frac{\alpha(Q^2)}{\pi} \right) ^3
- 130 \left( \frac{\alpha(Q^2)}{\pi} \right) ^4 ]\:(I_3+I_8)  +
\nonumber  \\
 + \frac{1}{6} [ 1 -  \frac{\alpha(Q^2)}{3 \pi}
 - 0.549 \left( \frac{\alpha(Q^2)}{\pi} \right) ^2
- 2 \left( \frac{\alpha(Q^2)}{\pi} \right) ^3  ]\:
 \frac{2}{3} \Sigma_{inv}.
\label{qcd2}
\end{eqnarray}

\vspace{3mm}

In the following we shall discuss the influence of the flavor singlet
evolution function on the sum rule in eq.(\ref{QCD}) and how
the flavor singlet amplitude $I_{0,model}$, eq.(\ref{I0}) (or $\Sigma_{model}$)
calculated in
the usual spectroscopic quark models which include
quark confinement, should be included in the two  expressions,
eqs.(\ref{QCD}) and (\ref{qcd2}).
Two extreme viewpoints regarding the influence
of the flavor-singlet evolution function  $Y(\alpha(\mu^2),\alpha(Q^2))$
 \cite{Jaffe,JaffeMa90} and \cite{GluckReya} were discussed, and still
is a matter of debate. One viewpoint argues that it is
not inconceivable that non-perturbative
effects could cause an evolution down to the very low
spectroscopic scale $\mu^2$
such that $Y(\alpha(\mu^2),\alpha(Q^2))$ is a very small number, a number
which in itself could explain the EMC result \cite{Jaffe,JaffeMa90}.
This scenario implies  that the
confinement quark model calculation predicts the matrix element of
the flavour singlet axial current $I_0(\mu^2)$ in eq.(\ref{QCD}).
Furthermore, such
an argument would lead to practically no
constraint on a quark model calculation of $I_0$.
In one of our previous works \cite{HHFM2} we adopted
this viewpoint, see e.g.  Refs.\cite{Jaffe,Mulders}, and discussed the effects
the known behaviour of the renormalization group exponent
could have on the "proton spin content".
This contrasts with our original attitude \cite{HHFM1}:
using models with confined quarks to compute
the matrix elements appropriate for the Bjorken
limit,  we compute  physical observable quantities, quantities which
should be independent
of the renormalization point and therefore should be identified
with the renormalization group invariant quantities \cite{GluckReya}
or $\Sigma_{inv}$ of eq.(\ref{sigmainv}).
This means only perturbative QCD corrections outlined in eq.(\ref{qcd2})
should be included in the sum rule.
The reason being that
any realistic quark confinement model (including chiral bag models) has
built-in long distance effects, meaning soft
gluonic corrections are to some extent
included in the quark wave functions of these models.
In addition, the chiral bag model also includes the "very  soft"
effective mesonic mediators of the longest range nuclear forces.
These effective $\qbarq$ $0^-$ degrees of freedom are important for
understanding properties of baryons.
This is of course very different from the parton model approach where  in the
first appoximation the valence quarks are free non-interacting quarks
with no gluonic content.

\vspace{3mm}

After these discussions
we can present some numerical estimates of what is expected
for the term $Y(\alpha(\mu^2),\alpha(Q^2))$ $I_0(\mu^2)$
of eq.(\ref{QCD}) and compare with
the  $\Sigma_{model}$ value calculated in the chiral bag model
where the pseudoscalar mesons are excluded from the bag.
We use the values for $F$ and $D$ of the
chiral bag model calculated in the chiral limit   \cite{HHFM2}
including the two-body magnetic gluon exchange spin-spin correction:
$F$ = 0.455 and $D$ = 0.795 ($F/D$ = 0.57).
This means that in eq.(\ref{QCD}) $I_3$ + $I_8$ = $F$ + $D$/3 = 0.72.
As  values for $\alpha(Q^2)$  we use the ones corresponding to
$\Lambda$=240 MeV  \cite{Alt94}
that runs within one standard deviation of data from the $Z_0$-mass to
the $\tau$-mass.
Then we find using the experimental values of eq.(\ref{one}):
\footnote{
If we use the "word average" value
\begin{eqnarray}
Y(\alpha(\mu^2),\alpha(Q^2))
I_0(\mu^2) =  0.213  \pm 0.053  \pm 0.073.  \nonumber
\label{main2}
\end{eqnarray}}
\begin{eqnarray}
Y(\alpha(\mu^2),\alpha(Q^2)) I_0(\mu^2) =  0.194  \pm 0.073  \pm 0.073.
\label{main}
\end{eqnarray}

\noindent
This result should be compared with the direct calculation of  the
chiral bag, which gave $I_{0,model}$ = 0.22 \cite{HHFM1},
corresponding to a "spin content" of
$\Sigma_{model}$ = 0.33 when the pseudoscalar mesons were excluded from the
interior of the bag.
\footnote{ In a version of the model where the pseudoscalar field
is continuous through the bag surface (cloudy bag),
$I_{0,cloudy}$ = 0.38 corresponding to $\Sigma_{cloudy}$ = 0.57 \cite{HHFM1},
the same value as in the Ellis and Jaffe sum rule.}
Early in the spin-crises period \cite{EMC1}, this
calculated value, $\Sigma_{model}$ = 0.33,
was considered to be too large,  which certainly,
with the new experimental results \cite{SMC} and the refined
calculations of the radiative corrections, no longer
is the case.
The latest Ellis and Karliner \cite{EllisKa94}
analysis gives  $\Sigma(Q^2=10 GeV^2) = 0.31 \pm 0.07$
(when higher twist effects are ignored).
Since $I_{0,model}$ is very close the the present value of
eq.(\ref{main}), there is in this model
little room for the scenario with $Y$ being very small.
However, as discussed above, since the quark wave function of
the chiral bag model contains soft-gluon effects, we should and shall identify
$\Sigma_{model}$ = 0.33 with $\Sigma_{inv}$ of eq.(\ref{qcd2})
when we calculate $\Gamma_p(Q^2)$.
The corresponding result for $\Gamma_n(Q^2)$ is obtained from the above by
substituting $I_8 - I_3$ = -0.53 for $I_8 + I_3$ = 0.72 in the same
equation (\ref{qcd2}).
In figures 1 and 2 we show both the model results using eq.(\ref{qcd2})
and the experimental data
as a  function of $\alpha(Q^2)$. We have not included any
corrections from higher twists that would influence the results at low $Q^2$,
the reason being that there is doubt both about the  size as well as the sign
of these corrections \cite{twist1}.
As we can see, the model is giving results that  are very satisfactory
when compared with
the experiments on both the proton and the neutron targets.

\vspace{3mm}

In this model the so-called
proton spin content has many components.  One third comes
from the spin of the three valence quarks, the remainder originates
from the orbital angular momentum  (of the relativistic quarks)
and from the gluonic exchanges between the valence quarks which can
be regarded as confined $ \qbarq$ pairs \cite{HHFM}.
Note that the
U(1) anomaly is essential in reconciling the correct values of F and D with
a value of $\Delta \Sigma < 0.57$.
In terms of the $ \Delta q$'s\ the model gives $\Delta u=0.83$, $\Delta
d=-0.42$
and $\Delta s=-0.08$ as $Q^2 \rightarrow \infty$.

\vspace{3mm}

We have also noted that a recent calculation by Narison, Shore and Veneziano
based on QCD spectral sum rules give results \cite{Veneziano} that are
very similar to the ones we have presented.
Our results were calculated in the chiral limit for the three quarks
where $SU(3)_f$ by
definition is a good symmetry group. When quarks and
pseudoscalar mesons are given masses
($m_u = m_d \approx 10$ MeV and $m_s \approx 200$ MeV),
$SU(3)_f$ is broken, then  what are the changes?
The neutral pion is fairly light but both $\eta$ and $\eta '$
become heavy. Due to their heavy
masses, the influence of both of the two
pseudoscalar $\ssbar$ carrying mesons are then suppressed.
This leaves almost no hidden strangeness around the
nucleons, as shown in an earlier
calculation for nucleons \cite{HHFM1}.
However, the results
for $\Gamma_{p,n}(\infty)$ hardly change at all!, meaning
the  chiral bag model with broken $SU(3)_f$ symmetry
and almost no $\ssbar$ content can describe data as well as the
 $SU(3)_f$ symmetric chiral bag model above.
{}From these results it follows that the experimental data of  today
do not necessarily imply that $\Delta s \neq 0$ but
but could just as well mean that $SU(3)_f$  symmetry is broken.
This specific example should underline the fact that the extraction of
$\Delta s$ from the data on the basis of $SU(3)_f$ symmetry should be taken
with some caution.

\vspace{3mm}

When a model is able to reproduce pre-EMC data on static properties
of the baryons such as the axial charge, coupling strength of pions to the
baryons, magnetic moments etc., and also give reasonable fits to
the SMC data \cite{SMC} in a new domain when the
effect of the U(1) anomaly on the $\eta^{\prime}$ mass is taken into account,
there is a coherence that give us confidence that we have some
understanding of
the rudimentary aspects of the nucleon structure.

\medskip

Correspondence with E. Reya are gratefully acknowledged.
This work is supported in part by NSF grant no. PHYS-9310124,
a NATO travel grant and a grant from Norges Forskningsr\aa d.

\section{Figure Caption}
\noindent
\begin{description}
\item[{\rm Figure 1:}]
Comparison of the chiral bag model calculation, eq.(\ref{qcd2}),
with experimental data of $\Gamma_p(Q^2)$.
Systematic errors are heavy vertical lines, statistical errors are thin
vertical
lines. Data are plotted for reasonable values of $\alpha(Q^2)$.
\end{description}
\noindent
\begin{description}
\item[{\rm Figure 2:}]
Same as Figure 1 but for $\Gamma_n(Q^2)$.
\end{description}

\end{document}